\documentclass[useAMS,usenatbib]{mn2e}
\usepackage{graphicx}
\bibpunct{(}{)}{;}{a}{}{,}

\title[Optimum frequency band for radio polarisation observations]
   {Optimum frequency band for radio polarisation observations}

\author[T.G. Arshakian et al.]
  {Tigran G.~Arshakian\thanks{On leave from the Byurakan
  Astrophysical Observatory, Aragatsotn prov. 378433, Armenia, and
  Isaac Newton Institute of Chile, Armenian Branch} and Rainer Beck \\
  Max-Planck-Institut f\"ur Radioastronomie, Auf dem H\"ugel 69, 53121 Bonn, Germany}

\date{Released 2011 Xxxxx XX}

\pagerange{\pageref{firstpage}--\pageref{lastpage}} \pubyear{2002}

\def\LaTeX{L\kern-.36em\raise.3ex\hbox{a}\kern-.15em
    T\kern-.1667em\lower.7ex\hbox{E}\kern-.125emX}

\begin{document}

\label{firstpage}

\maketitle

\begin{abstract}
Polarised radio synchrotron emission from interstellar, intracluster
and intergalactic magnetic fields is affected by frequency-dependent
Faraday depolarisation. The maximum polarised intensity depends on
the physical properties of the depolarising medium. New-generation
radio telescopes like LOFAR, SKA and its precursors need a wide
range of frequencies to cover the full range of objects. The optimum
frequency of maximum polarised intensity (PI) is computed for the
cases of depolarisation in magneto-ionic media by regular magnetic
fields (differential Faraday rotation) or by turbulent magnetic
fields (internal or external Faraday dispersion), assuming that the
Faraday spectrum of the medium is dominated by one component or that
the medium is turbulent. Polarised emission from bright galaxy
disks, spiral arms and cores of galaxy clusters are best observed
at wavelengths below a few centimeters (at frequencies beyond
about 10~GHz), halos of galaxies and clusters around
decimeter wavelengths (at frequencies below about 2~GHz).
Intergalactic filaments need observations at meter wavelengths (frequencies below 300~MHz). Sources with extremely large
intrinsic $|RM|$ or RM dispersion can be searched with mm-wave
telescopes. Measurement of the PI spectrum allows us to derive the
average Faraday rotation measure $|RM|$ or the Faraday dispersion
within the source, as demonstrated for the case of the spiral
galaxy NGC~6946. Periodic fluctuations in PI at low frequencies are
a signature of differential Faraday rotation. Internal and external
Faraday dispersion can be distinguished by the different slopes of
the PI spectrum at low frequencies. A wide band around the
optimum frequency is important to distinguish between varieties of
depolarisation effects.
\end{abstract}



\section{Introduction}

The major radio continuum surveys planned with future radio
facilities like the Square Kilometre Array (SKA), its precursor
telescopes ASKAP, MeerKAT and APERTIF, and low-frequency radio
telescopes such as LOFAR and MWA will open a new era in the
study of cosmic magnetic fields via polarised synchrotron emission
and Faraday rotation. As these telescopes will operate at different
frequencies, it is crucial to investigate which astrophysical
objects can be observed with a certain telescope and to select
the frequency band that will yield maximum polarised intensity for
these objects.

In total radio continuum intensity, many astrophysical sources
reveal a power-law synchrotron spectrum with an almost constant
spectral index over the radio frequency range where the energy
losses of the cosmic-ray electrons are small. The total intensity of
synchrotron emission depends on the number density of cosmic-ray
electrons and the strength of the total magnetic field component
normal to the line-of-sight of the observer, while the polarised
intensity is related to ordered magnetic fields. Ordered fields can
be regular (coherent), generated by the mean-field dynamo (Beck et
al. 1996) or anisotropic, generated from turbulent magnetic fields
by compressing or shearing gas flows. Turbulent fields with random
orientations give rise to unpolarised synchrotron emission. The
degree of synchrotron polarisation is a function of the ratio
between ordered and turbulent fields (Sokoloff et al. 1998).

If the magnetic field structure is not resolved, the degree of
polarisation is reduced by an effect called {\em beam
depolarisation} which depends on the beamsize of the
telescope. For the same resolution, the intensity of polarised
radio continuum emission is the result of competition between two
processes: synchrotron emission and {\em Faraday depolarisation}\
(DP), both of which increase with wavelength.

DP is caused by variations of Faraday rotation. Faraday rotation
changes the polarisation plane when the radio wave passes through a
magneto-ionic medium with regular magnetic fields. Hence, Faraday
rotation is an important signature of magneto-ionic media containing
regular magnetic fields and is a measure of field strength and
thermal electron density.

Faraday rotation $\Delta\chi$ is traditionally measured from the
polarisation angles $\chi$ at several wavelengths and quantified by
the rotation measure (RM), defined as $\Delta\chi = RM \, \Delta
\lambda^2$. The $\pm n\,\pi$ ambiguity of the polarisation angle
$\chi$ requires the determination of RM by the slope of the
best fit of the relation between $\chi$ and $\lambda^2$ -- if this
relation is linear.

Faraday rotation in a foreground screen in front of the
synchrotron-emitting region can be described by a single RM value
which means that the slope of the relation between $\chi$ and
$\lambda^2$ is constant over the whole wavelength range. If,
however, Faraday rotation occurs within the emitting region, the
observable RM is no longer constant beyond a critical wavelength
(Burn 1966): the medium becomes ``Faraday-thick''. Below this
critical wavelength, a ``simple'' layer can still be characterized
by a single value of RM, if the distributions of regular magnetic
fields and thermal electrons are box-like (``Burn's slab'') or
symmetric Gaussians or symmetric exponentials (Sokoloff et al.
1998).

In complex media with several distinct synchrotron-emitting and
Faraday-rotating regions within the measured volume, no single RM
value exists and {\em RM Synthesis}\ needs to be applied. It
Fourier-transforms the complex polarisation (amplitude and angle)
measured over a large frequency spread into the complex Faraday
spectrum in Faraday depth space (Burn 1966, Brentjens \& de Bruyn
2005, Heald 2009, Frick et al. 2010). Modern radio telescopes have
a sufficiently large number of frequency channels and large total
bandwidth to perform RM Synthesis with high resolution in Faraday
space.

Present-day data from the Westerbork Synthesis Radio Telescope
(WSRT) towards bright regions in the Milky Way and towards galaxy
clusters indicate that a significant (possibly dominant) fraction of
Faraday spectra show one component or one dominant component
(Schnitzeler et al. 2009, Pizzo et al. 2011, Brentjens 2011). Media
with turbulent magnetic fields and/or turbulent distribution of
thermal electrons are expected to show a turbulent Faraday spectrum
(Frick et al. 2011).

If the region contains cosmic-ray electrons, thermal electrons and
regular magnetic fields, the polarisation planes from waves from the
far side of the emitting layer are more Faraday-rotated than those
from the near side. This leads to wavelength-dependent
depolarisation, called {\em differential Faraday rotation} (DFR).
Turbulent fields also cause wavelength-dependent depolarisation,
called {\em Faraday dispersion}. Internal Faraday dispersion (IFD)
occurs in an emitting and Faraday-rotating region, while external
Faraday dispersion (EFD) may occur in a non-emitting foreground
screen (Burn 1966, Sokoloff et al. 1998). DFR is a function of RM
and wavelength (Eq.~(\ref{eq:dfr})), while Faraday dispersion
depends on RM dispersion and wavelength (Eqs.~(\ref{eq:ifd}) and
(\ref{eq:efd})).

Depolarisation of the emission from various cosmic objects varies
strongly and depends on coherence length, strength of the regular
and turbulent magnetic fields and thermal electron density. Hence,
each population of polarised objects should be studied at the
optimum wavelength at which the PI is maximum.

The observed spectrum of PI is a power law over a limited wavelength range and often reveals a maximum at a certain
wavelength $\lambda_{max}$ (Kronberg et al. 1972, Conway et al.
1974, Tabara \& Inoue 1980). Below $\lambda_{max}$, the degree of
polarisation decreases with decreasing wavelength, called {\em
polarisation inversion}. In the case of compact radio sources,
polarisation inversion is often related to flat-spectrum (opaque)
sources and is probably caused by Faraday dispersion (Conway et al.
1974).

In this paper, we investigate the optimum frequency band for
polarisation observations of various classes of astrophysical
objects. We assume that the Faraday spectrum is dominated by one
component or that the medium is turbulent. We also explore
possibilities of distinguishing between internal/external
Faraday dispersion and differential Faraday rotation, which allows
investigation of the physical properties of the depolarising
medium in various cosmic objects.


\section{Optimum wavelength for maximum polarised intensity}

\label{sec:dep} The total intensity of the synchrotron emission
detected in the rest frame of the observer at a frequency $\nu$ is
\begin{equation}
  I_{\nu} = C_1 \, n_{\rm CR} \, B_{\rm t,\,\bot}^{2(1+\alpha)} \, \nu^{-\alpha} \, L ,
  \label{eq:Inu}
\end{equation}
where $n_{\rm CR}$ is the density of cosmic-ray electrons (per
energy interval) which have a power-law energy spectrum ($N(E)
\propto E^{-\gamma}$) with the spectral index $\gamma$, leading to
the synchrotron spectral index $\alpha=(\gamma-1)/2$. $L$ is the
linear size of the emitting region, and $B_{\rm t,\,\bot}$ is the
strength of the total magnetic field perpendicular to the line of
sight.

The PI is given by
\begin{equation}
  P_{\nu} = p_0 \, I_{\nu} \, \left(\frac{B_{\rm ord,\,\bot}}{B_{\rm t,\,\bot}}\right)^2  \, DP_{\nu} ,
    \label{eq:Pnu}
\end{equation}
where $p_0=(1+\gamma)/(7/3+\gamma)$ is the maximum degree of
polarisation ($p_0\simeq0.74$ for a typical spectral index of
$\gamma\simeq2.7$ in galaxies), $B_{\rm ord,\,\bot}$ is the strength
of the ordered (regular + anisotropic \footnote{An anisotropic field
can be generated by compressing or shearing an isotropic turbulent
field; it contributes to polarised emission but not to Faraday
rotation.}) magnetic field perpendicular to the line of sight and
$DP_{\nu}$ is the depolarisation coefficient. Assuming that the
cosmic-ray density, total and ordered magnetic fields are
stationary, we write
\begin{equation}
    P_{\nu} = C \, \nu^{-\alpha} \, DP_{\nu},
    \label{eq:Pnus}
\end{equation}
where $C = C_1 \, p_0 \, n_{\rm CR} \, B_{\rm ord,\,\bot}^{2} \,
B_{\rm t,\,\bot}^{\alpha-1}$.


\subsection{Differential Faraday rotation}
\label{subsec:dfr}

Wavelength-dependent Faraday depolarisation occurs in a region
containing cosmic-ray electrons, thermal electrons and regular
magnetic fields. The polarisation planes of the waves from different
synchrotron-emitting layers are rotated differently: the
polarisation planes from the near emitting layers rotate less than
those emitted from the far layers. This effect is known as
DFR and is given by
\begin{equation}
  DP = \frac{|\sin(2 \, RM \, \lambda^2)|} {|2 \, RM \, \lambda^2|} .
  \label{eq:dfr}
\end{equation}
RM is the average observed rotation measure (in radians per square
meter),
\begin{eqnarray}
  RM\, [\mbox{rad m}^{-2}]= 0.81 \int n_e \, B_{\rm reg,\,\|} \, dL
        \nonumber \\
  \simeq 0.81 \, \langle n_e \rangle \, \langle B_{\rm reg,\,\|} \rangle \, L ,
  \label{eq:rm}
\end{eqnarray}
where $n_e$ (in $cm^{-3}$) is the thermal electron density, $B_{\rm
reg,\,\|}$ (in $\mu$G) is the strength of the regular magnetic field
along the line of sight, and $L$ is the pathlength through the
regular field and thermal gas in parsecs ($pc$). We assume here that
the magneto-ionic medium can be characterized by one single RM
value, i.e. the distributions of $n_e$ and $B_{\rm reg,\,\|}$ are
smooth and symmetric along the line of sight (Sokoloff et al. 1998).

\begin{figure}
    \center
    \includegraphics[angle=-90,width=8cm]{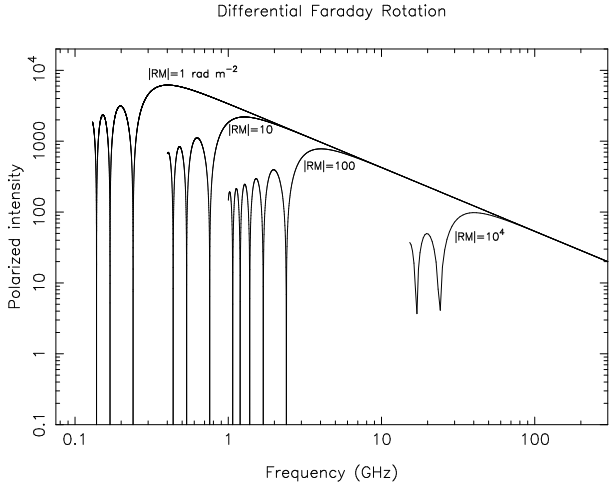}
    \caption{Polarised intensity for a spectral index of
    total synchrotron intensity of $\alpha=0.9$
    and depolarisation by differential Faraday rotation at the level
    of $|RM|=1$, $10$, $100$, and $10^4$~rad~m$^{-2}$. }
    \label{fig:dfr}
\end{figure}

\begin{figure}
\begin{center}
    \includegraphics[angle=-90,width=8cm]{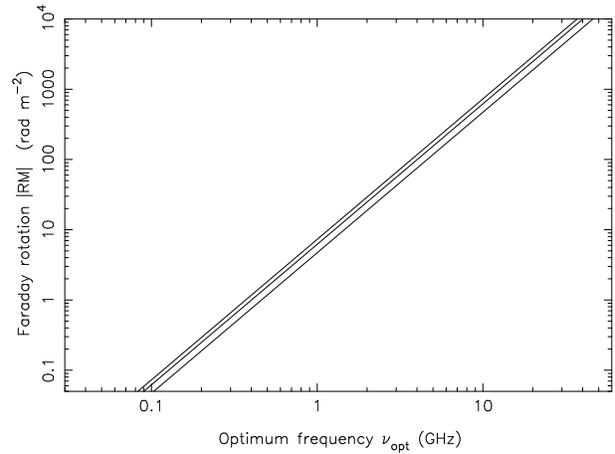}
\caption{Optimum frequency of maximum polarised emission for a
synchrotron spectrum with spectral index $\alpha=0.5$, $0.9$ and
$1.3$ (from bottom to top) and depolarisation by differential
Faraday rotation as a function of $|RM|$. } \label{fig:lambda-rm}
\end{center}
\end{figure}

The maximum PI is reached at larger frequencies for
larger values of RM (Fig.~\ref{fig:dfr}). At low frequencies (before
the maximum) the slope of the curve, measured between 1/10 and 1/100
of the maximum PI, is $\alpha \simeq100$
(Fig.~\ref{fig:dfr}). The periodic changes of $DP$ with wavelength
(Eq.~(\ref{eq:dfr})) lead to total depolarisation at certain
wavelengths, observable as ``depolarisation canals'' in maps of
polarised emission (e.g. Fletcher \& Shukurov 2006). However,
``canals'' can also originate from steep gradients in polarisation
angle caused e.g. by turbulent fields (Sun et al. 2011).

Accounting for the differential Faraday depolarisation
(Eq.~(\ref{eq:dfr})) and solving the equation $dP_{\nu}/d\lambda =
0$, we derive a transcendental equation for the optimum wavelength
($\lambda_{\rm opt}$) of the maximum polarised emission,
\begin{equation}
    |\sin k| - \frac{2k}{2-\alpha}|\cos k| = 0 ,
\end{equation}
where $k = 2 \, |RM| \, \lambda_{\rm opt}^2$. We derive the equation
\begin{equation}
\lambda_{\rm opt} = A(\alpha) \, |RM|^{-0.5},
\end{equation}
where $\lambda_{\rm opt}$ is measured in $m$ and $A(\alpha)$ is
0.65, 0.75, 0.81 for $\alpha=0.5, 0.9, 1.3$. The dependence of the
optimum frequency ($\nu_{\rm opt}$) on rotation measure for
$\alpha=0.5$, 0.9 and 1.3 is shown in Fig.~\ref{fig:lambda-rm}:
polarised sources with larger $|RM|$ are best observed at high
frequencies.

Note that regions with thermal electrons and a constant regular
field, but without cosmic-ray electrons (no synchrotron emission),
called ``Faraday screens'', cause Faraday rotation of polarised
emission from background sources, but do not depolarise. Any
variation of strength or direction of the regular field within the
volume traced by the telescope beam causes RM gradients and hence
depolarisation (Burn 1966, Sokoloff et al. 1998) which is similar to
external Faraday dispersion (see below).

\begin{figure}
\begin{center}
    \includegraphics[angle=-90,width=8cm]{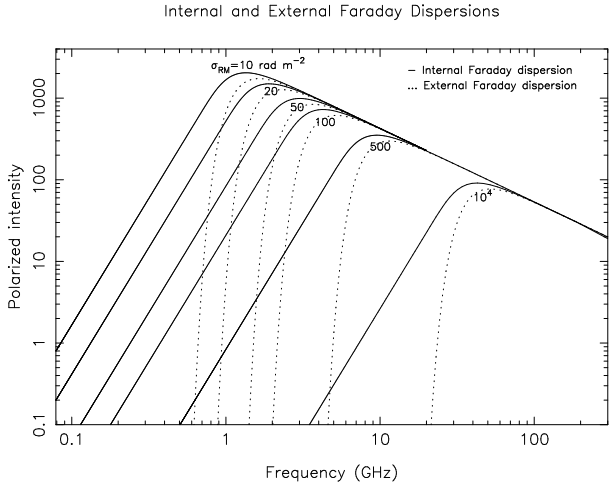}
\caption{Polarised intensity for a spectral index of
    total synchrotron intensity of $\alpha=0.9$
    and depolarisation by internal (solid line) and external
    (dashed line) Faraday dispersions at different levels of
    $\sigma_{\rm RM}$. } \label{fig:ifr_log}
\end{center}
\end{figure}

\begin{figure}
\begin{center}
    \includegraphics[angle=-90,width=8cm]{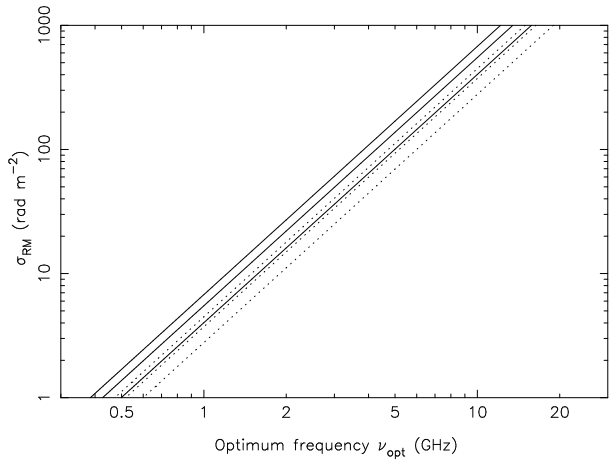}
\caption{Optimum frequency of maximum polarised emission for a
synchrotron spectrum with spectral index $\alpha=0.5$, $0.9$ and
$1.3$ (from bottom to top) and depolarised by internal (solid line)
and external (dashed line) Faraday dispersions against RM
dispersion.} \label{fig:lambda_sigrm}
\end{center}
\end{figure}

\subsection{Faraday dispersion}
\label{subsec:fd}

Depolarisation by IFD occurs in a region containing cosmic-ray
electrons, thermal electrons and turbulent magnetic fields and is
given by
\begin{equation}
  DP = \frac{1- e^{-S}}{S} ,
  \label{eq:ifd}
\end{equation}
where $S = 2 \, \sigma_{\rm RM}^2 \, \lambda^4$.
Depolarisation by EFD in a non-emitting Faraday screen is given by
\begin{equation}
  DP = e^{-S}.
  \label{eq:efd}
\end{equation}
The RM dispersion can be described in a simplified model of a
turbulent magneto-ionic medium as
\begin{eqnarray}
\sigma_{\rm RM}^2&=(0.81 \,n_e\, B_{\rm turb}\, d)^2 \, f \, L / d
        \nonumber \\
  &\cong (0.81 \, \langle n_e \rangle \, \langle B_{\rm turb} \rangle)^2 \, L \,d / f ,
\label{eq:sigma_RM}
\end{eqnarray}
where $n_e$ (in $cm^{-3}$) is the electron density within the
turbulent cells of size $d$ (the ``correlation length'', in $pc$),
$L$ is the pathlength (in $pc$), $\langle n_e \rangle$ is the
average electron density in the volume along the pathlength
traced by the telescope beam, $f$ is the filling factor of the
cells ($f=\langle n_e \rangle/n_e$) and $\langle B_{\rm turb}
\rangle$ (in $\mu$G) is the mean strength of the turbulent magnetic
field, assumed to be the same inside and outside of the cells. We
further assume that the field direction is constant within each cell
and that the contribution of the beamsize to $\sigma_{\rm RM}$
is negligible. If however the beamsize corresponds to a scale much
larger than that of RM variations, $\sigma_{\rm RM}$ cannot be
described by Eq.~(\ref{eq:sigma_RM}).


Note that other definitions of $\sigma_{\rm RM}$ in the literature
used a different dependence on the filling factor $f$. Future
high-resolution radio observations are needed which can directly
measure $\sigma_{\rm RM}$.

The effect of depolarisation of the PI by internal and external
Faraday dispersions is shown in Fig.~\ref{fig:ifr_log} for
$\alpha=0.9$. The dependence of polarisation intensity on wavelength
is the same for both mechanisms at high frequencies where no
depolarisation occurs, while beyond the peak (at low frequencies)
the PI decreases faster in the case of external Faraday dispersion.
At low frequencies (before the maximum) the slopes of internal and
external Faraday dispersion curves are significantly different: the
spectrum is a power-law ($P_{\nu} \propto \nu^{4-\alpha}$) for
internal Faraday dispersion, while for external Faraday dispersion
it deviates from a power law. The slope of the latter is estimated
to be $\alpha \simeq 15$ between 1/10 and 1/100 of the maximum PI.
The intensity fixed at the level of $\sigma_{\rm RM}$ reaches the
peak at a slightly larger frequency for external Faraday dispersion
than in the case of internal Faraday dispersion
(Fig.~\ref{fig:ifr_log}).

The equation for the optimum wavelength ($\lambda_{\rm opt}$) of
maximum polarised emission in the case of internal RM dispersion is
\begin{equation}
   e^{2S_{\rm o}}-\frac{8 \, S_{\rm o}}{\alpha-4}-1=0 ,
  \label{eq:sig_alpha_int}
\end{equation}
where $S_{\rm o}=2\sigma_{\rm RM}^2 \lambda_{\rm opt}^4$. For
external RM dispersion we derive the equation
\begin{equation}
   \lambda_{\rm opt}=\left(\frac{\alpha}{8 \, \sigma_{\rm RM}^2}\right)^{1/4}
   ,
  \label{eq:sig_alpha_ext}
\end{equation}
where $\lambda_{\rm opt}$ is measured in $m$.

The dependence of the optimum frequency on internal dispersion
(full line) and external (dotted line) dispersion is shown in
Fig.~\ref{fig:lambda_sigrm} for $\alpha=0.5, 0.9$ and $1.3$.
Polarised sources with larger $\sigma_{\rm RM}$ are best observed at
high frequencies. In the case of internal RM dispersion we
found that $\lambda_{\rm opt}=A_1\, \sigma_{\rm RM}^{-0.5}$, where
$A_1=0.6,0.7,0.87$ for $\alpha=0.5,0.9,1.3$. For external RM
dispersion the relations are $\lambda_{\rm opt}=A_2\, \sigma_{\rm
RM}^{-0.5}$, where $A_2=0.50,0.58,0.63$ for $\alpha=0.5,0.9,1.3$.

At long wavelength and/or large Faraday dispersion ($S>>1$)
Eq.~(\ref{eq:efd}) can no longer be applied because the correlation
length of polarised emission is smaller than the cell size $d$
(Tribble 1991, Sokoloff et al. 1998), and the external
depolarisation by external dispersion becomes
\begin{equation}
  DP = (2 \,\sigma_{RM} \, \lambda^2)^{-1}.
  \label{eq:efd2}
\end{equation}
This equation is valid only at wavelengths much longer than the
optimum wavelength which corresponds to $S_{\rm o}=\alpha/4<1$ (see
Eq.~\ref{eq:sig_alpha_ext}) and hence is not relevant for this
paper.

\subsection{Mixed cases}

Many astrophysical media contain both regular and turbulent magnetic
fields, while the descriptions of Faraday depolarisation in Sections
2.1 and 2.2 are only valid if one type of magnetic fields dominates.
In mixed cases with similar field strengths the total depolarisation
can still be described by Eq.~(\ref{eq:ifd}) where $S$ becomes a
complex number (Sokoloff et al. 1998). As an approximation, it may
be assumed that some fraction of the emitting medium on the far side
is totally depolarised by Faraday dispersion and the remaining
volume on the near side is subject to depolarisation by differential
Faraday rotation. Here, the total depolarisation is the product of
Eqs.~(\ref{eq:dfr}) and (\ref{eq:ifd}) with appropriate weighting
according to the strengths of the regular and turbulent field
components.

\subsection{RM grids}

If RMs of a grid of bright, compact polarised sources behind
extended foreground objects are measured, the foreground media
act as Faraday screens and contribute to one single component in the
Faraday spectrum. Only foreground regions with significant polarised
emission may generate secondary peaks in the Faraday spectrum. The
main depolarisation mechanism for RM grids is EFR in the foreground
(see Sect.~3). DFR and IFD may occur in the background sources, but
are generally weak due to the small source sizes and are further
reduced in distant objects by the RM dilution factor (see below).

\section{Discussion and conclusions}

An observer planning polarisation observations needs to
investigate the expected range of $|RM|$ and Faraday dispersion
within a source. In Table~1 we compiled typical physical properties
of magneto-ionic media in various astrophysical objects. The numbers
may vary by a factor of several or are still uncertain, as in the
case of the intracluster medium in galaxy clusters and of the
intergalactic medium. The media are assumed to be ``simple'',
characterised by a single RM component and/or by a RM dispersion
$\sigma_{\rm RM}$. The resulting optimum frequency bands give
a first-order estimate for the range of highest polarised
intensities.

Table~1 allows an observer to estimate which depolarisation
effect dominates in a medium: the larger the optimum
frequency, the stronger the depolarisation. In disks and halos of
galaxies, DFR and IFD are of similar importance. In ``magnetic
arms'' between optical spiral arms, the regular field and hence DFR
are strongest. In galaxy clusters, turbulent fields and hence IFD
dominate.

The polarised emission of the inner disks, spiral arms, central
regions of galaxies and the cores of galaxy clusters should be
observed at wavelengths below a few centimeters (at frequencies
beyond about 10~GHz), in order to avoid strong depolarisation by DFR
and IFD. Outer galaxy disks, galaxy halos, halos of galaxy clusters
and intergalactic filaments have lower intrinsic $|RM|$ and Faraday
dispersion and are best observed at decimeter wavelengths (at
frequencies below about 2~GHz). Polarised intensity from
intergalactic filaments is low because the predicted magnetic fields
are weak, but, due to the synchrotron spectrum, increases towards
the meterwave range where Faraday depolarisation is still small.
Observations with low-frequency telescopes such as LOFAR are
promising, but difficult due to the strong Galactic foreground.

The observed Faraday dispersion in the nearby ISM of the Milky Way
is about 10~rad~m$^{-2}$ at high latitudes and beyond
50~rad~m$^{-2}$ at low latitudes (Schnitzeler 2010), in agreement
with the model by Sun \& Reich (2009). EFD in the Galactic
foreground with an RM dispersion of 60--160~rad~m$^{-2}$ at low
Galactic latitudes (Sun \& Reich 2009) yields an optimum observation
wavelength of 5--7~cm, while about 20~cm (1.5~GHz) at high Galactic
latitudes. The all-sky RM surveys with the SKA (Gaensler et al.
2004) and its pathfinder ASKAP (project {\em POSSUM}; Gaensler et
al. 2010) and deep RM grids towards nearby galaxies and galaxy
clusters with the MeerKAT and APERTIF telescopes are planned around
1~GHz. At the low frequencies observed with LOFAR, the polarised
emission of the Galactic foreground is affected by Faraday rotation
and strongly fluctuates with position and frequency, which hampers
the detection of signals from extragalactic objects.

Sources with extremely large intrinsic RM ($|RM|\ga
10^4$~rad~m$^{-2}$) or large RM dispersion ($\sigma_{\rm
RM}\ga10^4$~rad~m$^{-2}$) are rare. Extreme rotation measures are
measured for core-dominated quasars, e.g.
3C\,273 ($\approx +4\times 10^4$ rad m$^{-2}$ over the range
43--86~GHz; Attridge et al. 2005), and for the Galactic Center,
Sgr\,A* ($\approx -5\times 10^5$ rad m$^{-2}$ over the range
150--400~GHz; Macquart et al. 2006, Marrone et al. 2007). Sources
with such large intrinsic RM will escape detection in polarisation
with the upcoming radio surveys (e.g. {\em POSSUM}) because of
strong depolarisation around 1.4~GHz (see Figs.~\ref{fig:dfr} and
\ref{fig:ifr_log}). The optimum frequency band of such objects is
beyond about 30~GHz and they can be targeted by $mm$-wave telescopes
operating at sufficiently high frequencies (e.g. ALMA). The
innermost regions of jets of core-dominated quasars, cores of
massive galaxy clusters and starburst galaxies having dense
turbulent gas and strong magnetic fields are candidates for such
extreme values of RM or $\sigma_{\rm RM}$.

The optimum wavelength band to observe distant polarised
sources is larger than for nearby ones. The intrinsic $|RM|$ and
intrinsic RM dispersion of distant objects observed at a fixed
frequency are smaller by a factor of $(1+z)^{-2}$ and, hence,
Faraday depolarisation is smaller and the degree of polarisation is
higher (Fig.~\ref{fig:dfr}). The optimum wavelength band to observe
a nearby bright galaxy disk with $|RM|(z=0)=200$~rad~m$^{-2}$ is
around $5$~cm (6~GHz, see Table~\ref{tab:1}). If we want to
observe the same source, for example, at $z=3$ then the observed
$|RM|(z=3)\simeq12$~rad~m$^{-2}$ and the optimum wavelength 
band for observation is around 20\,cm (1.4~GHz). The optimum
wavelength band to observe a cluster core with $\sigma_{\rm
RM}(z=0)\simeq1000$~rad~m$^{-2}$ (Table~\ref{tab:1}) at $z=3$ is
around 7~cm (4~GHz). Hence, cluster cores and bright disk
galaxies at high redshifts can be detected with ASKAP and the
high-frequency SKA array.

Most of distant sources detected with future sensitive radio
telescopes will be from the population of star-forming galaxies
which can be expected to be more polarised at larger distance where
depolarisation is smaller.
On the other hand, the detection of regular magnetic fields via RM
from intervening galaxies on the line-of-sight towards polarised
background sources (Bernet et al. 2008, Kronberg et al. 2008) will
become more difficult for distant galaxies because their $|RM|$ is
lower.


\begin{figure}
\begin{center}
    \includegraphics[angle=-90,width=8cm]{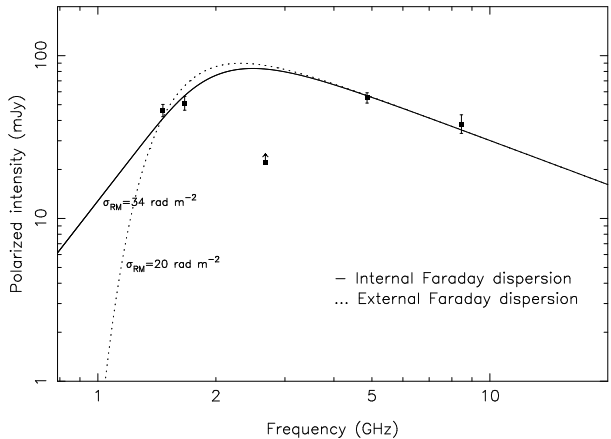}
\caption{Spectrum of integrated polarised flux densities of the
spiral galaxy NGC~6946 and fits with IFD and EFD. The data are
computed from the maps by Beck (2007) with the same resolution of
15'', except for the map centered at 2.675~GHz with 4.4'
resolution which yields a lower limit.} \label{fig:n6946}
\end{center}
\end{figure}

Depolarisation in galaxies can be well described by Faraday
dispersion. An example for the spiral galaxy NGC~6946 is shown in
Fig.~\ref{fig:n6946}. The spectrum of integrated polarised flux
densities is fitted by an IFD model (Eq.~(\ref{eq:ifd})) with
$\sigma_{\rm RM}\simeq34$~rad~m$^{-2}$, in excellent agreement with
the analysis of the depolarisation map by Beck (2007), or by an EFD
model (Eq.~(\ref{eq:efd})) with $\sigma_{\rm
RM}\simeq20$~rad~m$^{-2}$ from the Galactic foreground. With only
one more observation at a frequency band below 1.4~GHz, IFD
and EFD can be distinguished. Note that such integrated data cannot
be fitted by a DFR model because the variation in RM across the
galaxy smooths out the sharp minima seen in Fig.~1. DFR can only be
detected in objects with constant RM.

In this paper we demonstrated that measurement of the spectrum of PI
around the optimum frequency offers a simple first-order method to
measure the average $|RM|$ or the average Faraday dispersion in
media with a simple structure or strong turbulence, without
knowledge of polarisation angles.
Moreover, the knowledge of the optimum frequency and hence the main
(first) maximum of the PI spectrum is important for performing RM
Synthesis. It is presumed that a range around the optimum frequency
is included in the observed spectral range to ensure the recovery of
the main peak of the RM Transfer Function which is needed to clean
the Faraday spectrum (Heald 2009). Sufficiently wide coverage
in frequency is also needed to distinguish between varieties of
depolarisation effects.

We also showed that the slope of the PI spectrum at low frequencies
is much steeper for EFD than for IFD (Fig.~4). This allows us to
distinguish between these two effects, which is hardly possible with
other methods. DFR has a similarly steep spectrum as EFD, but is
easily recognizable by its periodic fluctuations, leading to total
depolarisation at certain wavelengths (Fig.~1).

If the synchrotron-emitting and Faraday-rotating medium has a
complicated structure but is not strongly turbulent, a spectrum of
components in Faraday space is expected, no well-defined peak of the
PI spectrum can be found and the results of this paper cannot be
applied. A model for two Faraday depth components was discussed by
Farnsworth et al. (2011). Two-component Faraday spectra have been
detected e.g. towards the inner regions of a few spiral galaxies
(Heald et al. 2009), possibly due to a different field configuration
in the nuclear region or a reversal of the radial field components
on the near and far sides of the nucleus. Faraday spectra towards
radio galaxies located in the inner parts of galaxy clusters also
reveal multiple components which may emerge from the lobes (Pizzo et
al. 2011). The fraction of multi-component Faraday spectra of ISM
regions in the Milky Way is still unclear. While most spectra
towards the Perseus cluster near to the Galactic plane are
complicated (Brentjens 2011), the ISM near the Galactic anti-centre
is Faraday-quiet (Schnitzeler et al. 2009). The forthcoming all-sky
RM survey {\em GMIMS} (Landecker 2010) will bring us more clarity.

\begin{table*}
\begin{minipage}[t]{18cm}
\caption{Typical properties of diffuse magneto-ionic media and the
corresponding optimum frequencies for polarisation observations,
assuming a synchrotron spectral index $\alpha=0.9$.}
\label{tab:1}              
\centering                          
\renewcommand{\footnoterule}{}  
\begin{tabular}{|l| c c c c c c c| c c | c c|}        
\hline\hline                 
{\bf Source} & $\langle n_{\rm e} \rangle $ & $B_{\rm reg}$ &
$B_{\rm turb}$ & $L$ & $d$ &$f$ & Ref. &$|RM|$~\footnote{Predicted;
consistent with observations where available}& $\nu_{\rm
opt}$~\footnote{Optimum frequency in case of differential Faraday
rotation} & $\sigma_{\rm RM}~^a$
& $\nu_{\rm opt}$~\footnote{Optimum frequency in case of internal Faraday dispersion} \\
           &  (cm$^{-3}$)           & ($\mu$G)        & $\mu$G)        & (pc)&  (pc) &  &
           & (rad m$^{-2}$) & (GHz) & (rad m$^{-2}$) & (GHz) \\ 
\hline                        
{\bf Emitting \& Faraday-} \\
{\bf rotating media} \\
\hline                        
Faint galaxy disk   & 0.01  & 5 & 5   & 1000 & 50  & 0.2 & 1  & 40  & 2.5 & 20  & 2.3  \\
Bright galaxy disk  & 0.05  & 5 & 10  & 1000 & 50  & 0.5 & 2,3& 200 & 6  & 130 & 6   \\
Spiral arm          & 0.1   & 2 & 20  & 500  & 50  & 0.5 & 3  & 80  & 3.8  & 360 & 10   \\
Magnetic arm        & 0.03  &10 & 5   & 500  & 50  & 0.5 & 2  & 120 & 4.3  & 27  & 2.7  \\
Star-forming complex& 0.5 & $<2$& 20  & 100  & 10  & 0.05& 3  &$<80$&$<3.8$&1100 & 15   \\
Faint galaxy halo   & 0.01  & 1 & 3   & 1000 & 50  & 0.5 & 4  & 8   & 1.2 & 8   & 1.5  \\
Bright galaxy halo  & 0.02  & 3 & 5   & 1000 & 50  & 0.5 & 4,5& 50  & 2.7 & 25  & 2.5  \\
Galaxy cluster halo & 0.001&$<1$& 1-5 &$10^5$&1-5~$10^4$&1?&6,7&$<80$&$<3.8$ & 25-300 & 2.5-10 \\
Galaxy cluster core & 0.01 &$<1$&10-30&$10^4$& 3000   &1? & 8 &$<80$ &$<3.8$ & 500-1500 & 10-30 \\
Galaxy cluster relic&$10^{-4}$&$<1$&1-5&$10^6$~\footnote{Projection
dependent} &
1000 & 1? & 9 & $<80~^{d}$ & $<3.8$ & 3-15 & 0.9-2  \\
IGM filament &$5~10^{-6}$&$<0.1$&$<0.3$&$5~10^6~^{d}$&$5~10^5$&1?&10&$<2~^{d}$
&$<0.6$~\footnote{Limited by the polarised emission of the Galactic foreground}&$<2$&$<0.75~^{e}$ \\
\hline                        
{\bf Faraday-rotating,} \\
{\bf non-emitting media } \\
{\bf (``Faraday screens'')} \\
\hline                                   
Local Milky Way~\footnote{High Galactic latitude} & 0.03  & 2 & 3  & 200 & 50  & 0.5 & 11 & 10 & 1.3 & 10 & 1.7  \\
Local Milky Way~\footnote{Low Galactic latitude}  & 0.05  & 2 & 5  & 3000& 50  & 0.5 & 11 & 80 & 3.8  & 80 & 5   \\
IGM around radio lobes          &0.001& $<1$& 1-5  &$10^5$&5-20~$10^3$& 1? & 12,13&$<80$&$<3.8$&20-200 & 2.3-7.5 \\
\hline                                   
\end{tabular}
\end{minipage}

\medskip
\begin{flushleft}
{\bf References} 1: Fletcher et al. (2004), 2: Beck (2007), 3:
Fletcher et al. (2011), 4: Hummel et al. (1991), 5: Heesen et al.
(2009), 6: Kim et al. (1990), 7: Murgia et al. (2004), 8: Vogt \&
En{\ss}lin (2005), 9: van Weeren et al. (2010), 10: Xu et al.
(2006), 11: Sun \& Reich (2009), 12: Laing et al. (2008), 13: Feain
et al. (2009).
\end{flushleft}

\end{table*}

%
%
%

\section*{Acknowledgments}
This work was supported by the European Community Framework
Programme 6, Square Kilometre Array Design Study (SKADS). TGA
acknowledges support by DFG--SPP project under grant 566960. We
thank Luigina Feretti and Torsten En{\ss}lin for help in compiling
cluster data for Table~1, as well as Roberto Pizzo, Wolfgang Reich,
Dominic Schnitzeler, Rodion Stepanov, and Dmitry Sokoloff for useful
discussions.

\bibliographystyle{mn2e}

\label{lastpage}

\end{document}